# ESA F-Class Comet Interceptor: Trajectory Design to Intercept a Yet-to-be-discovered Comet.

Joan Pau Sánchez[a], David Morante[b], Pablo Hermosin[b], Daniel Ranuschio[a], Alvaro Estalella[a], Dayana Viera[a], Simone Centuori[b], Geraint Jones[c,d], Colin Snodgrass[e], Cecilia Tubiana[f]

[a]*School of Aerospace, Transport and Manufacturing, Cranfield University, Cranfield, UK*
[b]*Deimos Space S.L.U, Ronda de Poniente 19, Edificio Fiteni VI, Tres Cantos, Madrid, Spain*
[c]*Mullard Space Science Laboratory, University College London, UK*
[d]*Sentre for Planetary Sciences, UCL/Birkbeck, UK*
[e]*School of Physics & Astronomy, University of Edinburgh, UK*
[f]*Istituto di Astrofisica e Planetologia Spaziali – IAPS/INAF Roma, Italy*

**Abstract**

Comet Interceptor (Comet-I) was selected in June 2019 as the first ESA F-Class mission. In 2029[+], Comet-I will hitch a ride to a Sun-Earth L2 quasi-halo orbit, as a co-passenger of ESA's M4 ARIEL mission. It will then remain idle at the L2 point until the right departure conditions are met to intercept a yet-to-be-discovered long period comet (or interstellar body). The fact that Comet-I target is thus unidentified becomes a key aspect of the trajectory and mission design. The paper first analyses the long period comet population and concludes that 2 to 3 feasible targets a year should be expected. Yet, Comet-I will only be able to access some of these, depending mostly on the angular distance between the Earth and the closest nodal point to the Earth's orbit radius. A preliminary analysis of the transfer trajectories has been performed to assess the trade-off between the accessible region and the transfer time for a given spacecraft design, including a fully chemical, a fully electric and a hybrid propulsion system. The different Earth escape options also play a paramount role to enhance Comet-I capability to reach possible long period comet targets. Particularly, Earth-leading intercept configurations have the potential to benefit the most from lunar swing-by departures. Finally, a preliminary Monte Carlo analysis shows that Comet-I has a 95 to 99% likelihood of successfully visit a pristine newly-discovered long period comet in less than 6 years of mission timespan.

## 1. Introduction

While the scientific return of past comet missions is unquestioned, all previously visited comets had approached the Sun on many occasions and, consequently, have also undergone substantial compositional and morphological modifications. Comet Interceptor mission proposes instead to intercept and explore an undiscovered Long Period Comet (LPC), which ideally would be visiting the inner Solar System for the first time.

The European Space Agency (ESA) selected in 2019 the Comet Interceptor (Comet-I) as its first Fast-track class mission (F-class). This newly created F-Class mission category enables mission opportunities that exploit joint launch opportunities with other ESA missions, and emphasises implementation of novel space concepts. Comet-I will be launched in 2029 as a co-passenger with ESA's M4 ARIEL Mission on board an Ariane 6.2. It will be deployed in a Sun-Earth L2 (SE L2) quasi-halo orbit and remain there by means of small orbit maintenance manoeuvres for a period of up to 3 years; while it awaits for the right departure conditions to intercept a yet-to-be-discovered long period comet.

After selection, Comet-I successfully completed the internal ESA studies of the Assessment Phase (phase 0) [1]. In February 2020, ESA approved advancement of Comet-I into the Definition Phase (Phase A). The current baseline involves three spacecraft elements working together to ensure a bountiful scientific return through unprecedented multipoint measurements. A main spacecraft (S/C A) would make remote observations of the target from afar, to protect it from the dust environment and act as the primary communications hub for all other mission elements. Two smaller daughtercraft (JAXA's S/C B1 and ESA's S/C B2) would venture much closer to the target, carrying instruments to complement and enhance the scientific return.

A key aspect of Comet-I is the need to be designed without an identified target, or rather Comet-I must be designed to intercept a yet-to-be-discovered comet. Hence, a sufficiently large number of comet intersection trajectories must be optimized and studied, in order to inform the spacecraft design with statistical underpinning. Moreover, the trajectory analysis must also bring evidences that Comet-I is not subject to the seemingly chanceful event of an adequate LPC being discovered, but that, on the contrary, the risk for Comet-I not having an adequate fly-by opportunity is indeed very low.

The present paper synthesises the work carried out at Cranfield University and DEIMOS Space to understand and inform the feasibility of Comet-I mission design. The paper presents the highlights of three parallel 5 to 6 months MSc internship projects carried out during the second and third quarters of 2020 [2–4].

## 2. Trajectory Design Requirements

This first section examines the main set of requirements and constraints that must be satisfied by Comet-I's intercept trajectory. These are derived from two main sources: On the one hand, the science objectives of the mission, which obviously require Comet-I to reach a pristine newly-discovered comet. On the other hand, the boundary conditions set for the F-Class Mission, which in fact define some of the most constraining limitations on the trajectory design, such as the feasible $\Delta v$ budget.

### 2.1. Scientific Requirements

Comet-I's primary goal is to provide the first-ever characterization of a pristine newly-discovered long period comet (LPC) [5]. The comet characterisation must include morphological analysis, surface composition, structure, composition of the coma, etc. The mission must also provide a multi-point measurement of the comet-solar wind interaction. Hence, given the necessary measurements, it follows that the mission must go far beyond what a typical Earth-based point source observation could achieve, and must indeed intercept with the targeted comet at a close range. Comet-I should encounter the comet within a distance of a few hundreds to a few thousands of kilometres, in order to allow for sufficient spatial resolution of the remote sensing payload [5].

Comet-I seeks to provide new understandings of the working of our Solar System, the processes of planetary formation and the emergence of life. To achieve this, it will visit a comet which has the Oort cloud as its point of origin. The Oort cloud is located at the very edge of our Solar System and provides a reservoir of the most pristine leftovers of the formation of planets. Ideally, the target comet will be visiting the inner solar system for the first time (generally referred as dynamically new comet). Finally, while unlikely and given their exceptional scientific value, there is a non-negligible probability that a serendipitous interstellar object, such as 1I Oumuamua, may be discovered and targeted by Comet-I [6].

Thus, feasible target comets will have orbital periods of tens of thousands of years (or hyperbolic trajectories), which implies that they will need to be first discovered by Earth-based astronomical surveys while in an inbound trajectory, and intercepted before it leaves the inner solar system. It follows then that the targeted comet is today undetected (i.e., still to be discovered) and its current position must be somewhere just past the orbit of Neptune (i.e., such that it reaches a periapsis after the $2029^+$ launch window).

Hence, the future target for Comet-I must be discovered sufficiently deep into the outer Solar System to allow sufficient warning time for a successful transfer to the intercept point. Until relatively recently these comets were discovered within Jupiter's orbit [7]. However, recent advances allowing deeper and more coherent surveys are lengthening the distance at which these objects are discovered. Moreover, the Vera C. Rubin Observatory's Legacy Survey of Space and Time (LSST) is scheduled to be online from the mid-2020s and will conduct the most sensitive Solar System survey ever performed [8].

*2.2. F-Class Boundary Conditions*

Comet-I is the first ESA Fast Track mission (F-Class). As quoted in the original Announcement of Opportunity1 (AoO), ESA's cost at Completion (CaC) must be less than 150 M€. However, this CaC does not include launch cost, since Comet-I will be launched towards the Sun-Earth Lagrange L2 point in a dual launch with ARIEL M4 mission. ARIEL is currently scheduled for launch in 2029 and, as such, Comet Interceptor must be launch ready by 2028 [9]. Hence, Comet-I must fit within the leftover launch performance of ARIEL's Ariane 6.2, which severely limits the available wet mass for Comet-I. At the time of the original AoO, the available wet mass was quoted as of 1,000 kg. However, after several design iterations, the latest requirement states that Comet-I should have no more 750 kg of wet mass, excluding launch adapter [1]. Given the CaC and the available wet mass, is clear that Comet-I should be a small mission/spacecraft and have a limited mission duration. In fact, the mission duration is expected to be of at most 6 years, including the transmission of all science data back to Earth.

---

[1]https://www.cosmos.esa.int/web/call-for-fast-mission-2018

*2.3. Trajectory Design Envelope*

The fact that Comet-I's target has neither been selected, nor discovered, means that the Δv cost to reach the target cannot be possibly known. More importantly, the available Δv budget will define the spacecraft's region of accessibility, which is to say the space within which intercept trajectories are feasible. Hence, given the epistemic uncertainties on the orbit of the yet-to-be-discovered target, the allocation of Δv budget has a direct relation with the probability of finding a feasible target for Comet-I. Here finding does not refer to the event of a pristine long period comet being newly surveyed, but to the probability of this comet having an interception point within reach of Comet-I spacecraft.

Following on Sec. 2.1 and with the purpose of providing a preliminary envelope for what can be considered feasible design options, this paper will assume a Δv ~ 1 km/s to be roughly feasible for the propulsive capabilities of Comet-I. It should be noted that Phase 0 mission requirements for ESA's Invitation to Tender quote a capability of 750 m/s as a requirement for the interception trajectory [1]. This is along the lines of the original Comet-I submission to the AoO, which also considers the trade-off between chemical and low-thrust propulsion system, and identified this as a key aspect to iterate during phase 0/A [2]. Indeed, this trade-off was preliminarily studied during Comet-I Phase 0, concluding that while 1.5 km/s may be a feasible capability for an electric propulsion system, the chemical propulsion configuration seems to cope better with the strict programmatic requirements of the F-Class mission [3]. Nevertheless, this will be subject to further iterations during Phase A/B1. Table 1 summarises Section 2 discussion on the trajectory requirements and constraints.

**Table 1. Comet Intercept Trajectory Requirements and Constraints.**

| # | *Trajectory Requirement and Constraints* |
|---|---|
| 1. | Comet-I must fly-by a pristine LPC within a few thousand kilometres from its nucleus. |
| 2. | LPC must be discovered sufficiently deep into the outer Solar System to allow the necessary time of flight to reach the intercept point. |
| 3. | Launch date in 2029. |
| 5. | Mission Duration less than 6 years. |
| 6. | Δ$v$ budget of about 1 km/s. |

## 3. The Population of Long Period Comets

As mentioned above, the goal of Comet-I is to understand the evolution of the Solar System by visiting one of its pristine building blocks, preserved during billions of years within the distant Oort cloud. These pristine building blocks are eventually nudged out of the Oort cloud, likely by the Galactic tide and nearby stars, and return to the inner Solar System as long period comets (LPC).

An LPC is generally defined as an object with orbital period larger than 200 years. If the object has reached the planetary region (i.e., q<10 au), means it will have a minimum eccentricity of about 0.7. However, as shown in Figure 1, there is a very strong bias towards eccentricities very nearly 1 (i.e., >0.99), as would be expected for objects with aphelion within the Oort cloud.

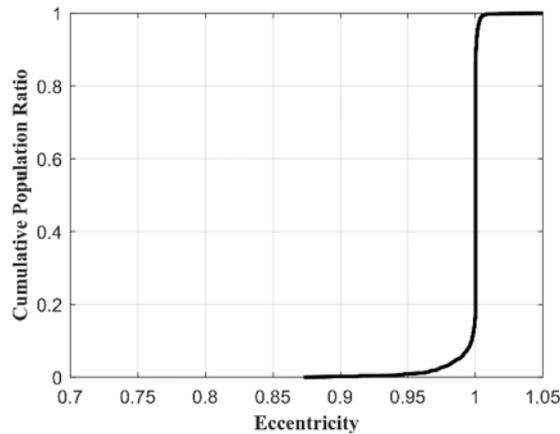

**Figure 1. Cumulative Probability Distribution of the eccentricity of the known population of LPCs. The eccentricity of 2668 LPCs was retrieved from JPL Small-Body Database Search Engine. Search includes all P>200 years, hyperbolic and parabolic comets with q<10 au.**

Figure 2 instead shows the probability density function of the periapsis distance for the same known population of LPCs (for the [0,2] au range). It can be seen that the known population have a clear bias at 0 au, however this bias is an observational artifact favouring the detection of sungrazing comets [4]. Thus, a better data set is the synthetic LPC population created by Boe et al [5] to simulate survey detection efficiency. The population was generated based on the best current knowledge of Oort cloud comet orbital distribution [6], and was shown to be a good match for the de-biased discovered population of the currently most sensitive survey, i.e. Pan-STARRS [5]. Boe et al [5] kindly provided the set of 1700 objects with q<2 au.

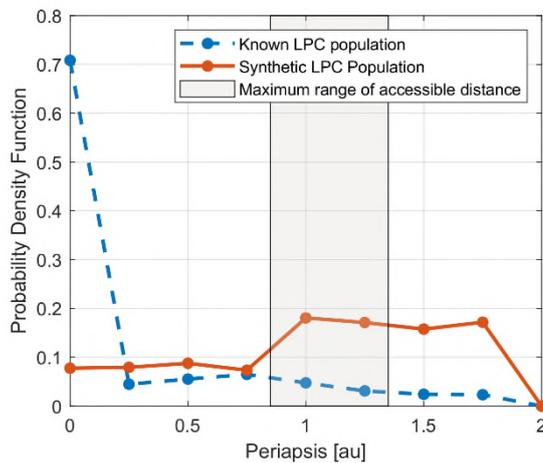

**Figure 2. Probability density function of LPC periapsis distance for the 0 to 2 au range for the known LPC population and a debiased synthetic LPC population by Boe et al. [5]. The range [0.85, 1.35] has been highlighted as potentially reachable by Comet-I spacecraft (arguably an optimistic case).**

While detailed discussion of Comet-I trajectory design will follow, it should now be noted that Comet-I limited $\Delta v$ budget will not allow to access to the whole heliocentric distance range, as plotted in Figure 2. However, the shadowed area is here assumed to be potentially accessible, at least if particular departure conditions benefiting from Moon swing-bys are achieved, as demonstrated by ISEE-3/ICE [7]. Within this range, and given the statistical significance of the plotted sample, it can be assumed that the probability density is relatively homogenous.

On the other hand, and given the cost of out-of-the-ecliptic transfers, it can also be assumed that Comet-I will most likely intersect the targeted LPC near the ecliptic. Figure 3 shows all the nodal points (i.e., ascending and descending nodes) of the known LPCs and Boe et al. synthetic population. As observed in the figure, the density of intersection points within the accessible region is approximately homogeneous as well. Finally, it should be noted that the decade 2010-2019 saw a total of 30 new LPCs with nodal points within the accessible range [0.85,1.35] au. This is an equivalent frequency of 3 LPCs a year, without consideration of the size of the comet.

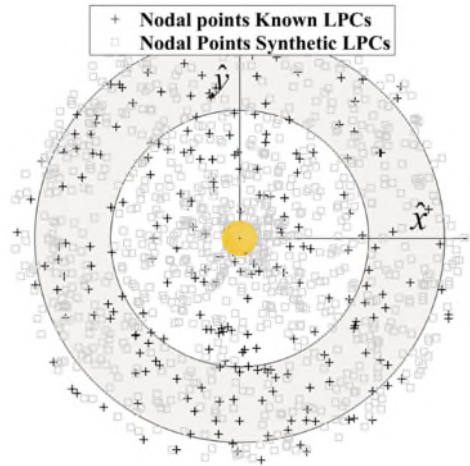

**Figure 3. Nodal point distribution in the ecliptic plane for all known LPCs and Boe et al. synthetic population [5].**

The actual size of the discovered comet will relate with its nuclear absolute magnitude, which, together with its coma activity, will define the heliocentric distance at which the comet will be discovered. The Comet should be discovered just about the orbit of Jupiter in order for Comet-I to have about 400 days of warning time. A warning time of 800 days would be equivalent to a discovery at 8.6 au distance and 1000 days of warning time would be equivalent to 10 au, approximately.

It is worth noting that state-of-the-art survey capabilities are currently detecting LPCs considerably beyond the orbit of Jupiter (see figure 5 from Królikowska and Dybczyński [8]). However, a key capability that may arguably enable Comet-I mission is the Vera C. Rubin Observatory's Legacy Survey of Space and Time (LSST), which is scheduled to be operational in 2023[1]. The LSST survey will scan the night sky at a visual limiting magnitude three points deeper than any previous survey. The survey simulation environment developed for LSST [9, 10] was used to simulate the detection efficiency of the synthetic LPCs generated by Boe et al. [5]. This simulation confirms that most of the LPCs discoveries are expected to have more than 3 years of warning time. In particular, 90% of the LPCs should be spotted beyond the orbit of Jupiter, 80% should allow for more than 2 years of warning time and about 75% should allow for more than 3 years of warning time [2].

Hence, overall, once the LSST is fully operational, 2 LPC per year should be expected to have intersection opportunities within the accessible region of the ecliptic plane (see grey area in Figure 3) and allow for at least 3 years warning time. While one more LPC may be spotted and only accessible with less than 3 years transfer time.

---

[1] https://www.lsst.org/about/timeline

## 4. Preliminary Trajectory Analysis

As previously discussed, Comet-I needs to be designed for an unknown target. Thus, maximizing the probability of having a successful mission, i.e., finding at least one accessible LPC given the $\Delta v$ budget and the available time-of-flight (ToF), is desirable. To get an overview of the accessible or reachable region for a given spacecraft design, a large number of intersection trajectories must be optimized and studied. In order to reduce the computational burden of this process and obtain representative results within a reasonable amount of time, two main assumptions have been made: a planar in-ecliptic trajectory and Two-Body dynamics.

Given the cost of performing out of ecliptic manoeuvres and that Perry [11] concludes that around 95% of the synthetic LPCs generated by Boe et al. [5] have their optimal point of interception at their ascending or descending node, the spacecraft dynamics has been reduced to a planar trajectory on the ecliptic plane. According to Perry's study [11], further considerations must be made if the radius of perihelion of the target is smaller than 0.1 au or the orbit inclination is low (i.e., smaller than 10º or larger than 170º), as the optimal interception location would be positioned close to the point of Minimum Orbital Intersection Distance. However, these corner cases were not considered in this paper, since they require a three-dimensional dynamical model that increases the computational cost and widens the search space.

Additionally, the dynamics is restricted to a perturbed Two-Body Problem (TBP) with the Sun as main body. Despite the fact that escape manoeuvres from SE L2 would be more accurately calculated using either the Circular Restricted Three-Body Problem or full ephemeris models, the impulsive modelling of the chemical propulsion system (CP) suggests that an approach based on the dynamics of a TBP will obtain similar results [12], while reducing the computing time. This assumption also holds for low-thrust electric propulsion (EP) transfers when the thrust-to-mass ratio is such that the burn time of each manoeuvre is small compared to the transfer duration [12]. A concession towards accounting for third-body effects has been made on the definition of the departure conditions. For the analysis carried out in this section, the orbit of the Earth is modelled as circular and the spacecraft is initially located at the SE L2 distance, orbiting the Sun with the same constant angular velocity of the Earth.

It should be noted that under the previous assumptions, any comet interception trajectory requiring an Earth leading configuration (i.e., *moving faster* than the Earth) requires the spacecraft to reduce its energy (i.e., semimajor axis), such that its period is shorter than that of the Earth. In a patched-conic framework, this will be equivalent to a relatively large $\Delta v$ manoeuvre, while in three-body dynamics this can be instead achieved at no $\Delta v$-cost, through a transit trajectory [13]. Since *free* connecting transfers in terms of delta-V exist between periodic orbits near the SE L2 and L1 points (see [14]), analysis assuming that the spacecraft starts the interplanetary transfer from SE L1 are also presented in this section.

The encounter point is fully defined by two parameters: Rc, which is the heliocentric radius of the comet's orbit when crossing the Ecliptic plane, and $\theta$, which is the phase angle at encounter, measured as Comet-Sun-Earth angle. The two dimensional or planar dynamics is expressed in terms of polar coordinates and the spacecraft is assumed to be controlled by a CP system, by an EP system or by a hybrid system that combines both. Notably, a full CP spacecraft has been already baselined in [12] as it relies on components with notable heritage to minimise the risk of the mission. However, the decision is, as of now, open for discussion. Therefore, and due to the fact that the use of CP limits the total applicable $\Delta v$ and consequently the accessible region, the possibility of using EP to transfer to the comet needs to be explored. Moreover, the exploitation of both propulsive system for the interplanetary transfer is studied in this work.

The interplanetary trajectory design problem is solved as an optimal control problem (OCP) where the goal is to find the control parameters that minimize the total $\Delta v$ budget (or the propellant mass) and comply with the mission requirements. The optimization variables are the magnitude and direction of the impulses for the CP trajectories, including one impulse at departure and one optional Deep Space Manoeuvre (DSM); the thrust steering law for the low-thrust EP trajectories and a combination of both for hybrid propulsion. A two-step algorithm is proposed as solution approach. Firstly, an analytical solution is obtained by computing the minimum $\Delta v$ Lambert arc that connects the departure point (either SE L2 or SE L1) with the given pair Rc-$\theta$. Thereafter, this solution is used as initial guess for a numerical optimizer that automatically transcribes the continuous OCP into a nonlinear programming problem (NLP). The transcription method is based on the Hermite-Simpson collocation approach and the interior-point solver IPOPT is employed as NLP solver. The interested reader is encouraged to check [15] and [16] where a similar approach is used to optimize the interplanetary transfer.

Applying this methodology allows to automatically and with minimal user-interaction perform a parametric search over the entire domain defined in terms of the encounter heliocentric distance Rc and phase angle $\theta$. It is worth mentioning that the same methodology works for the three propulsion systems under consideration. Results are produced in terms of reachability maps.

### 4.1. Chemical Propulsion Trajectories

The first strategy considers that the spacecraft performs an impulsive manoeuvre with CP to exit the Earth-Moon system. The spacecraft is instantaneously injected into a heliocentric orbit that starts drifting towards the location of the encounter with the target comet without performing any DSM. The maximum number of complete heliocentric revolutions is limited to be lower than 3. The optimizer solves for a grid of 350 points covering from 0.85 to 1.35 au. The performances in terms of required delta-V of the obtained points are linearly interpolated in a grid with 18,000 points. Results in the form of colour maps are provided in Figure 4, considering only the departure from SE L2 and without imposing any constraint on the time of flight. The accessible region is represented in purple colour, and include those trajectories that need less than 750 m/s to reach the comet, as derived from the requirements in [12]. Notice that the search space of the phase angle has been limited to -150º to +150º to avoid the solar conjunction and the loss of signal during the fly-by, according to the requirements defined in [12].

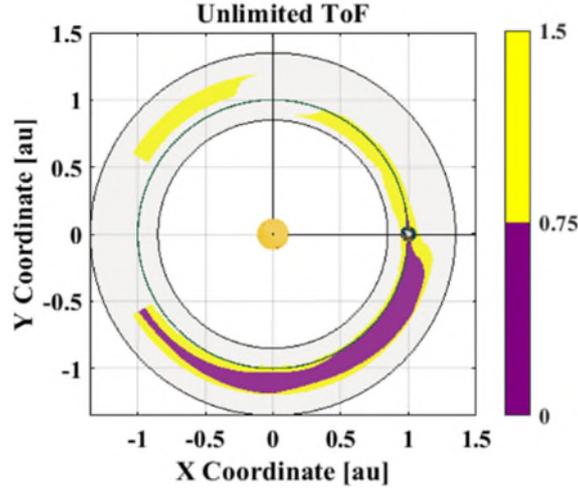
**Figure 4:** Accessible region for chemical propulsion departing from SE L2.

The plot reveals that departing from SE L2 without any DSM favours the intersection of Earth-trailing comets at a distance higher than 1 au. In fact, a comet interception ahead of Earth or below 1 au is non-feasible under these assumptions. This is due to the fact that departing from SE L2 implies a penalty of about 0.5 km/s for Rc < 1 au and a saving for Rc > 1.08 au [12]. The maximum accessible distance is 1.18 au, while the minimum distance is limited to 1 au. The maximum phase angle is found to be -150°. The longest transfer requires 1150 days and occurs for the farthest reachable point, i.e., Rc=1.10 au and $\theta$=150°. This region should be considered as underestimates of the real accessible area since, following previous discussion, given sufficient warning time, an effectively *free* transition between the L2 and L1 region should always be possible. Such a transition would likely enable accessible regions with a more symmetrical aspect in both trailing and leading regions (see further discussion on this point in Section 5).

In Figure 5, the accessible region including trajectories departing from SE L1 and from SE L2 are shown. Three different scenarios limiting the maximum transfer time to 400, 800, and 1000 days have been simulated. Note that in these plots, the constraint on the solar conjunction is not applied to have a complete overview of the feasible region. It can be observed that allowing trajectories departing from SE L1 makes Earth-leading trajectories with Rc < au 1 feasible. Thus, the possibility of finding a successful comet is increased. Notably, the feasible region is not completely symmetric w.r.t. Earth, being the maximum phase angle for Earth-leading trajectories lower than for Earth-trailing trajectories. This is due to the fact that reducing periapsis (i.e., going from SE L1 to Rc<1 au) has a larger cost than increasing the apoapsis (i.e., going from SE L2 to Rc>1 au) .

Additionally, Figure 5 showcase the effect of restricting the transfer time. In particular, reducing the transfer time does not modify the maximum accessible distance (i.e., 1.18 au) nor the minimum accessible distance (i.e., 0.86 au). However, reducing the transfer time significantly reduces the maximum phase angle for both Earth-leading and Earth-trading transfers. Notably, the farthest reachable point is always found at the same heliocentric distance, 1.10 au for Earth-trading trajectories, and 0.90 au for Earth-leading trajectories.

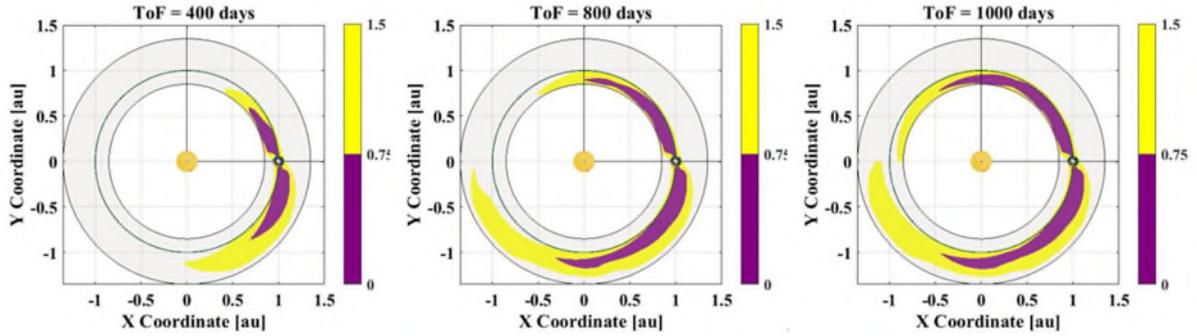
**Figure 5: Accessible regions for chemical propulsion. The color map represent the Δ$v$ budget in km/s**

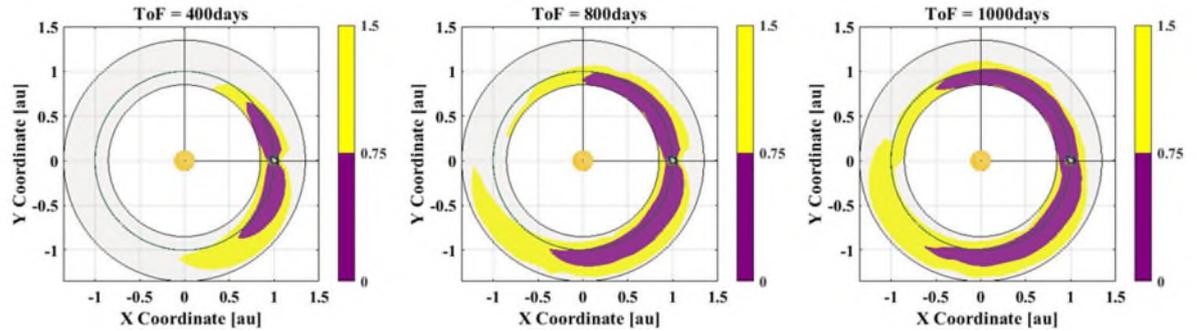
**Figure 6: Accessible regions for chemical propulsion with 1 DSM. The color map represent the Δ$v$ budget in km/s**

In Figure 6, a new set of simulations have been carried out, allowing a DSM on the way to the comet to adjust the orbit phasing when necessary. It can be seen that the accessible area has been extended to intersections with Earth-leading comets with Rc>1 au when departing from SEL 1. Furthermore, the new region also includes Earth-trailing comets with Rc<1 au when departing from SEL 2. In the former case, the DSM is used at the periapsis of the transfer orbit to increase its velocity and consequently the apoapsis, allowing the spacecraft to reach targets above 1 au. In the latter case, the DSM is applied in the apoapsis of the transfer orbit to reduce its velocity and consequently its periapsis, allowing the spacecraft to reach targets below 1 au.

### 4.2. Electric Propulsion Trajectories

This strategy considers that the spacecraft is instantaneously injected into a heliocentric orbit resulting from departing from the SE L2 or SE L1 initial conditions. Thereafter, the spacecraft is allowed to manoeuvre by means of EP, including thrusting and coasting arcs where necessary. Furthermore, the maximum number of complete heliocentric revolutions is limited to be lower than 3. The optimizer solves for a grid of 350 points covering from 0.85 to 1.35 au. The performances of the obtained points are linearly interpolated in a grid with 18,000 points. Results in the form of colour maps are provided in Figure 7. The accessible region is represented in purple colour, and include those trajectories that needs less than 1500 m/s to reach the comet, as derived from the requirements in [12]. Note that the colour map is expressed in terms of the propellant mass consumed. The limit for feasible trajectories is set to 72.74 kg as computed assuming an Isp of 1500s for the EP system and a spacecraft mass of 750 kg [12]. Two different EP thrusters has been simulated, one with 40mN and another one with 80 mN, in three different scenarios where the transfer time has been limited to 400, 800, and 1000 days.

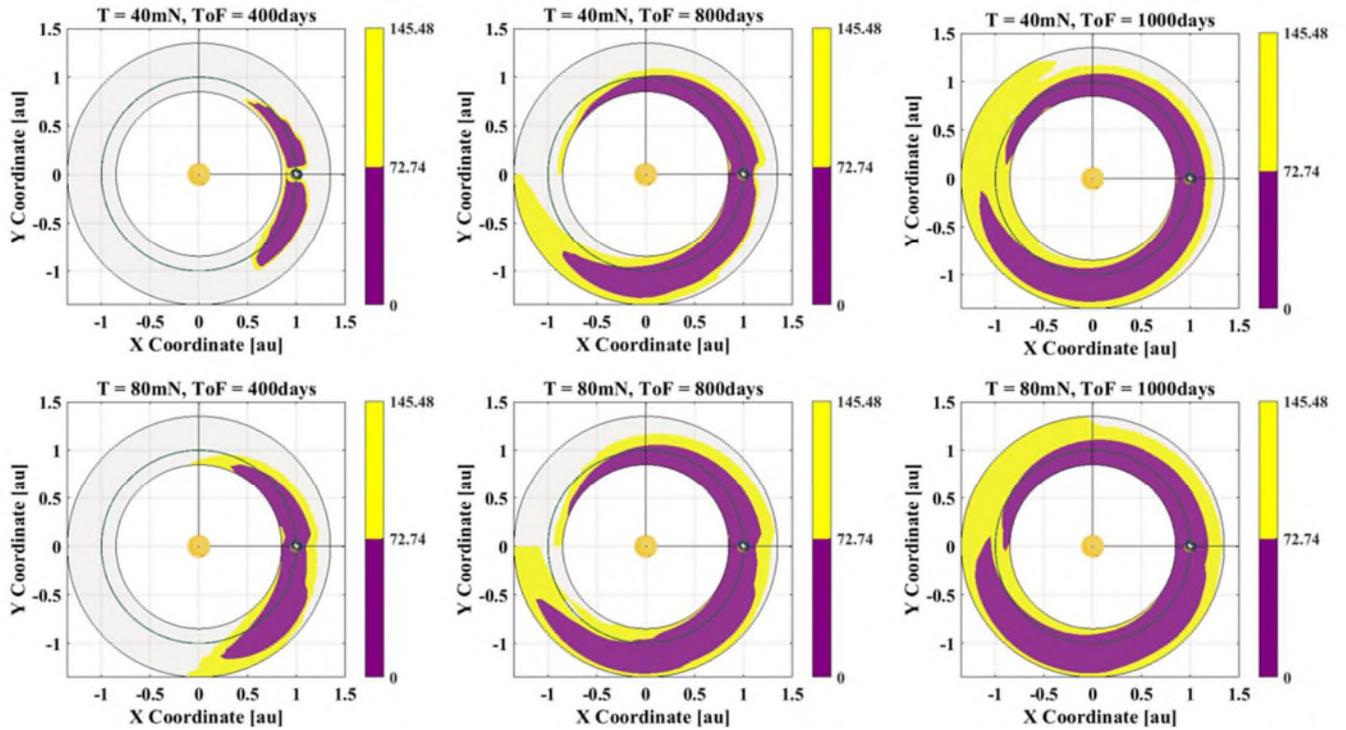

**Figure 7: Accessible regions for electric propulsion. The color map represent the propellant mass in kg**

Most notably, it can be deduced that by using EP as primary system for the interplanetary transfer the accessible region has been widen. In fact, when comparing to the results obtained for CP in Figure 6, it can be seen that the maximum phase angle and maximum distance are both increased. The yellow area in Figure 6 represents the reachable intersections for trajectories with a delta-V lower than 1500 m/s. This region can be seen as an upper boundary of the low-thrust feasible trajectories, i.e., the higher the thrust the closer the feasible region will be to the boundary. This is due to gravity losses resulting from the non-impulsive nature of the manoeuvres. Regarding the thrust/coast configuration the following results have been obtained. Earth-leading trajectories with Rc < 1 au and Earth-trading trajectories with Rc > 1 au consist on a thrust arc followed by coasting phase until intersecting with the comet. Earth-leading trajectories with Rc > 1 au and Earth-trading trajectories with Rc < 1 au includes an additional thrust arc at the periapsis and at the apoapsis respectively. These results match those conclusions derived from the analysis of the chemical trajectories with one DSM.

Furthermore, reducing the maximum available time or reducing the magnitude of the available thrust for the transfer reduces the maximum phase angle, both ahead and behind of Earth. In particular, when limiting the transfer time to 400 days, the spacecraft with 80 mN is able to reach comets with a phase angle of approx. 80° behind and ahead of Earth. However, if the spacecraft were propelled by 40 mN, it would only reach comets 55° behind and ahead Earth.

*4.3. Hybrid Propulsion Trajectories*

In this scenario, the spacecraft performs an impulsive manoeuvre with CP to depart from SE L2 or SE L1. The spacecraft is then instantaneously injected into a heliocentric orbit and is allowed to correct the trajectory by manoeuvring with an EP system. Following the same approach as in the previous subsections, a grid search has been performed for different levels of thrust, i.e., 40 mN and 80 mN, and for different maximum time-of-flights, i.e., 400 days, 800 days and 1000 days. The feasible region was defined in terms of propellant mass, the Isp of the EP was set 1500s, and the Isp of the CP was set to 300s.

The colour maps obtained are graphically similar to the results shown in Figure 7 for a full EP spacecraft. However, propellant mass savings are obtained. In Figure 8, a colour map representing the propellant mass savings in the feasible region comparing the hybrid system with 40 mN and 400 days maximum time of flight of the full EP system is shown. It can be seen that trajectories close to and far from Earth are the most benefited from the initial impulse, saving up to 60 kg of fuel.

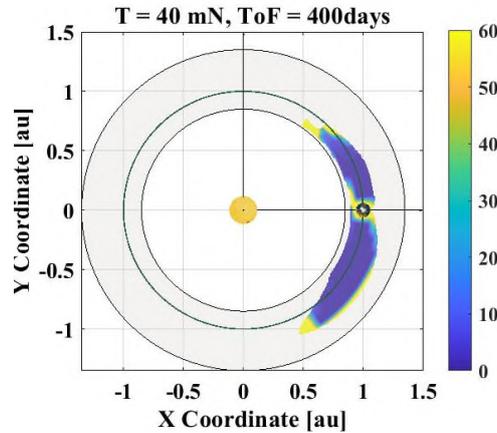

**Figure 8: Propellant mass savings (Kg) obtained with the Hybrid system compared to a full EP transfer.**

## 5. Earth-Moon System Departure Options

It should be reminded that the analysis in Section 4 assumes a dynamical model compatible with a patched-conic approximation. In a classic patch-conic framework, a departure trajectory is treated as a trajectory escaping the Earth's sphere of influence (SoI), and the Earth's SoI is assumed to be of negligible size (i.e., $R_{SoI}=0$), as compared with the Earth-Sun distance. However, similarly to Sánchez-Pérez et al. [12], here we have considered Comet-I as departing from the Sun-Earth L2 point, with the same angular velocity of the Earth. Such a departure condition would be equivalent to an excess velocity $v_\infty$ of about 300 m/s in a patched-conic framework. Note that here we compare the equivalent semimajor axis (see [17]), rather than the difference in velocity; thus the results differ slightly from those reported in [12].

As already pointed out in Section 4, *free* connecting transfers exist between periodic orbits near the SE L2 and L1 points (see [14]). Hence, if a new surveyed LPC is reachable by means of an Earth-leading transfer (i.e. transfer with a heliocentric orbit with a shorter than 1 year period), an effectively *free* transition between the L2 and L1 region should be possible, as long as sufficient warning/transfer time is available. Considering this, transfer in Section 4 assumed a convenient departure either from SE L2 or L1 point, in order not to over-penalise the Δ$v$ transfers costs due to a somewhat simplistic patched-conic dynamical framework. However, while a transfer from an L2 configuration to an equivalent L1 departure may be possible at a virtually negligible Δ$v$-cost, the total time of flight will not be negligible. This section will analyse in more detail Comet-I departure options. As a prelude of this analysis, it should be remarked that ISEE-3/ICE spacecraft [7] departed from a SE L1 Halo orbit to visit the Earth geomagnetic tail (i.e., L2 region), it then performed a double lunar swing-by and departed towards comet Giacobini-Zinner with an excess velocity $v_\infty$ of about 1.5 km/s.

*5.1. Exploiting the natural instability of the SE L2 point*

Both ARIEL and Comet-I spacecraft will be inserted by Ariane 6.2 into a free-insertion transfer to a Sun-Earth L2 large amplitude Halo orbit. While the exact orbit is not yet known, all available options (see [12]) represent departure conditions from higher energy levels than that of the specific L2 point; thus, potentially higher departure excess velocities too. However, an specific analysis of departure conditions is relevant given that all large amplitude halo orbit will have periods around half a year [18]. Thus, Comet-I may not have the opportunity to choose the optimal departure configuration from the Halo orbit. The analysis carried out here assumes a Halo orbit such as that in Figure 9; i.e., an 850,000 x 600,000 km $y$ and $z$ axis amplitude orbit, respectively.

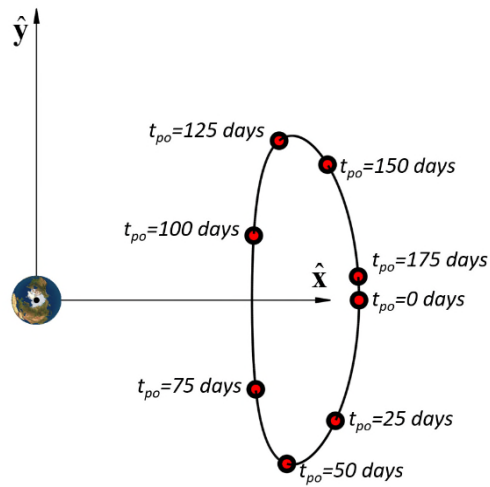

**Figure 9. Halo orbit view from positive z axis. The period of this Halo orbit is 179 days.**

Figure 10 shows the equivalent patched-conic excess velocity at departure (see [17]) as a function of different departing locations along the Halo orbit (i.e. $t_{po}$). The Halo orbit departure trajectories are here generated by perturbing the Halo orbit state vector in the direction of the unstable hyperbolic invariant manifold. Here, a Δ$v$ of 10 m/s is applied on the direction of the unstable eigenvector associated with the periodic orbit in order to generate the departure conditions. Given the asymptotic behaviour of the hyperbolic unstable manifold, smaller perturbations would result in slower departures from L2 [13], while larger Δ$v$ manoeuvres would result in lesser benefit on the form of *free* Δ$v$ [19].

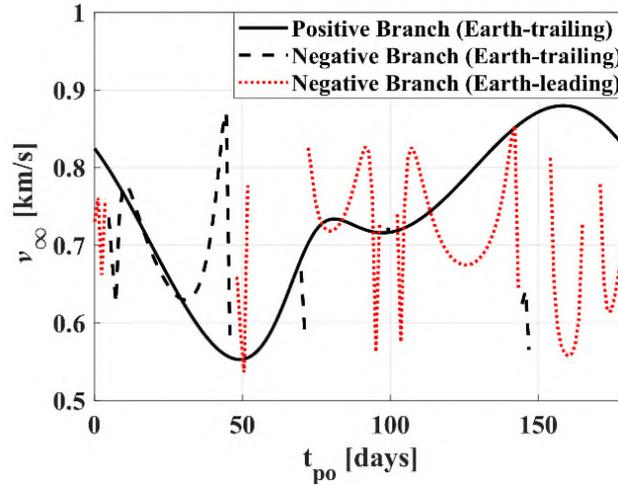
Figure 10. Escape velocity for direct departure trajectories

It should be noted that the hyperbolic invariant manifold has a positive branch that departs directly towards heliocentric trajectories, while its negative branch falls back into Earth centred trajectories. The trajectories in the negative branch can either finally depart through the L2 point towards heliocentric orbits with semimajor axis larger than 1 au and, consequently into Earth-trailing configurations, or depart through the L1 point into heliocentric orbits with semimajor axis smaller than 1 au (Earth-leading configurations). Indeed, it should be noted in Figure 10 that some data is unavailable for small particular ranges of departing conditions (i.e. $t_{po}$). This is due to the fact that their perturbed invariant manifold trajectories take too long to depart the Earth's neighbourhood to be considered in the analysis (> 2 years). The following subsection expand on the issue of the time of flight to complete the Earth departure phase.

*5.2. Combining manoeuvring and dynamical instability*

In order to speed up the escape trajectories through the unstable manifolds, some manoeuvres can be applied. For this study [19], direct escape trajectories from an orbit around SE L2 point (manifold positive branch) have been analysed. An initial impulse, of up to 0.2 km/s, in the direction of the unstable manifold eigenvector is applied at the departure points of the escape trajectory. The consequence of this non-negligible initial impulse is that escape times experience a great reduction (from around 250 to 95 days). Note that here the escape time is computed as the time for the trajectory to reach a distance to Earth of 0.02 au. The geocentric relative escape velocities ($v_\infty$) obtained for this group can be analysed on Figure 11, as a function of escape times and initial velocity impulse. It is noteworthy that the magnitude of the initial velocity impulse appears to be relatively ineffective at increasing the final escape velocities. The ranges of available escape velocity $v_\infty$ for at any given initial velocity impulse appear relatively unchanged and mostly affected by the departure point of the orbit. Instead, increasing the magnitude of the initial velocity impulse appears to be very effective at reducing transfer times. An example of these escape trajectory options can be observed in Figure 12.

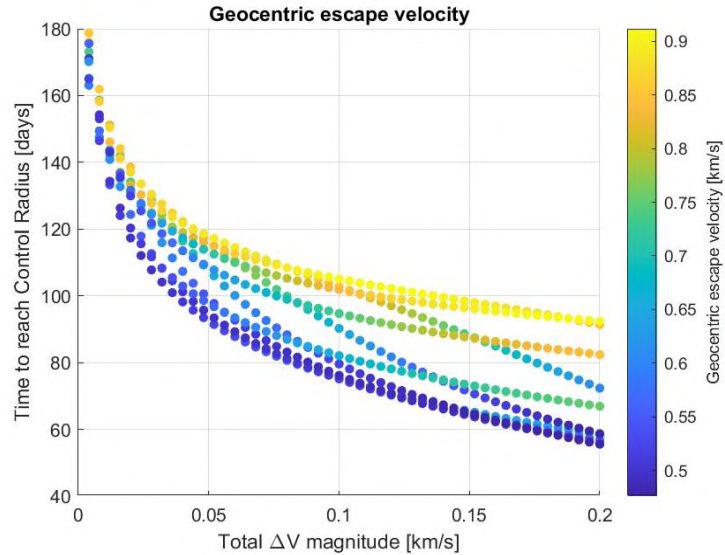

**Figure 11. Geocentric relative escape velocity $v_\infty$ for direct escape trajectory options.** Velocity increment versus escape time plot.

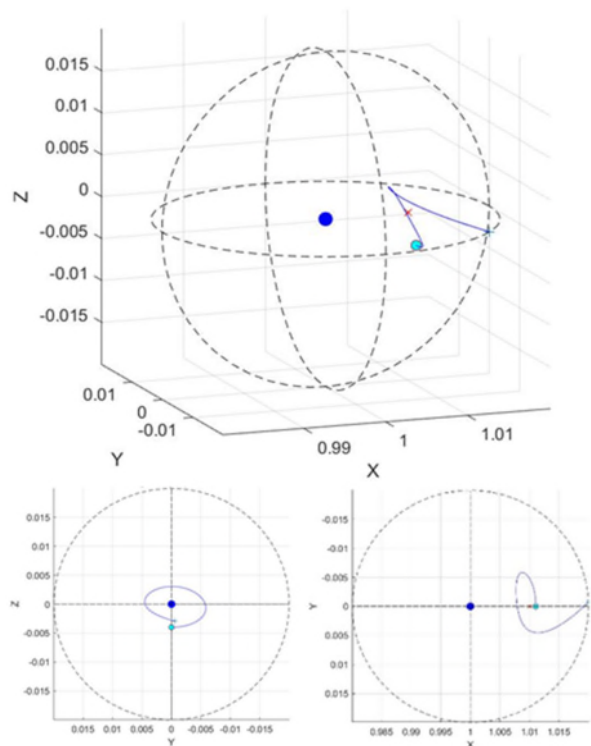

**Figure 12. Example of direct escape trajectory.** The blue line shows the trajectory. The Earth is plotted as a blue circle, on the centre of the control sphere, SE L2 point as a red cross. And the control sphere is illustrated in black dashed line. The image is displayed form three standpoints: perspective view, YZ plane and XY plane.

Trajectories that escape from an orbit around SE L2 point towards the Earth (manifold negative branch) are now added two manoeuvres to analyse the potential enhancement their performance. The manoeuvres include an initial impulse in the direction of the unstable manifold, similar to the one applied to the previous group, and a second impulse at the Earth's close approach. As shown in Figure 13, these trajectories also undergo a large decrease in escape time (from around 360 to 166 days). An added benefit of this type of departure option is that these trajectories also feature a non-negligible out-of-the-ecliptic component of the escape velocity, allowing flexibility to intercept the target comet slightly away from the ecliptic plane. Figure 14 displays an example of escape trajectory following the manifold negative branch.

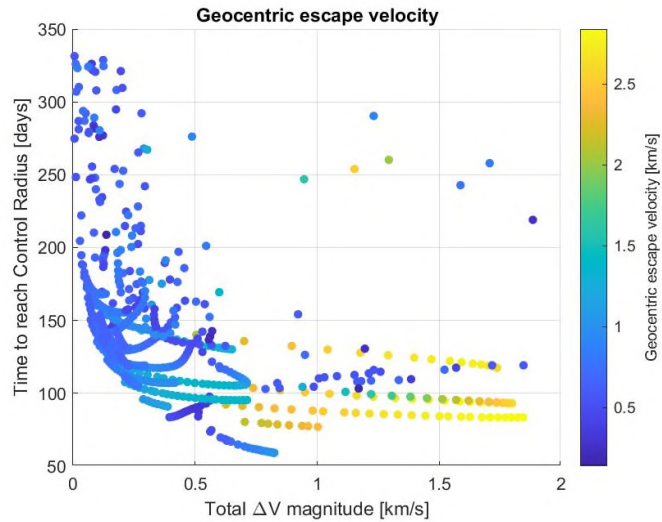

**Figure 13. Geocentric velocity, for escape trajectories with Earth approach.** Velocity increment versus escape time plot.

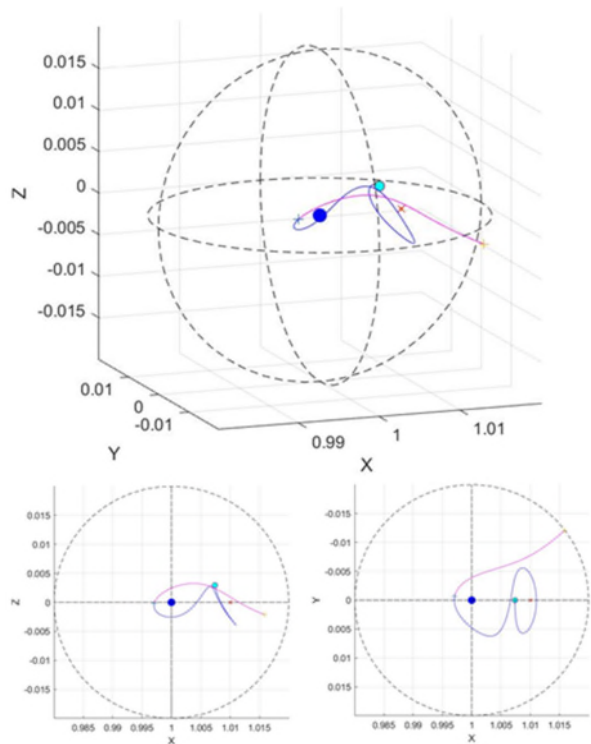

**Figure 14. Example of escape trajectory with Earth approach.** The blue line shows the first trajectory arc and the pink line shows the second one. The Earth is plotted as a blue circle, on the centre of the control sphere, SE L2 point as a red cross, and the departure point as a cyan small circle. And the control sphere is illustrated in black dashed line. The image is displayed form three standpoints: perspective view, XZ plane and XY plane.

Altogether, very efficient options to escape exist, depending on the available time and fuel. If the gain in geocentric escape velocity is prioritised, options that obtain up to 2.32 km/s of geocentric escape velocity can be found, which would need a velocity increment of 0.7 km/s and a time of flight of 104 days. Then, if the time to escape is considered to be the most important parameter, options needing 56 days are feasible, by applying an impulse of around 0.2 km/s, which reach geocentric escape velocities of approximately 0.5 km/s. On the other hand, if the main driver is purely the velocity increment cost, there are options from 0.004 km/s, that already offer moderate escape times (176 days) and geocentric escape velocities (0.86 km/s). It should be mentioned that the results of the obtained escape trajectory have a strong dependence on the orbit's departure point, as seen in Figure 10.

*5.3. Lunar swing-by departure options*

It is finally of relevance to note that most of the negative branch invariant manifold trajectories reach the orbit of the Moon. Hence, opening up the possibility of adding at least one lunar swing-bys to the departure phase. Figure 15 shows two example of Comet-I departure transfers, both following the negative branch of the hyperbolic manifold and including a single lunar swing-by. These trajectories were modelled as nearly-ballistic transfers in a Sun-Earth planar circular restricted three body problem (PCR3BP) with departure conditions defined within a planar Lyapunov orbit of similar *y* axis amplitude as the Halo orbit represented in Figure 9. The Moon flyby is here computed by means of a two-body patched conic framework, where the Moon flyby is modelled as an instantaneous rotation of the Moon-spacecraft relative velocity vector. Full details on the dynamical model can be found in Ranuschio [17].

In order to model departures trajectories such as those in Figure 15, the position of the spacecraft and that of the Moon along their respective orbits need to be known for the specific epoch of the discovery of the targeted LPC. Since the latter is not known and will not be known until the serendipitous discovery occurs, the initial ephemeris of both Moon and spacecraft need to be treated as homogeneously distributed random variables (due to their epistemic uncertainty). Hence, a Monte Carlo analysis is performed to understand the realistic expectations for the *free* $v_\infty$ excess velocity that can be gained through the Earth and Moon gravitational perturbations. Yet, given the sensitivities of the three-body motion near the Lagrange points, an optimization process is required for each Monte Carlo case to optimize the exact departure manoeuvre from the Libration Point Orbit, LPO, ($\Delta v \leq 10$ m/s), departure point along the LPO and altitude of the lunar fly-by that maximises the final $v_\infty$-equivalent excess velocity. Further details on the set up for this analysis can be read in Ranuschio [17].

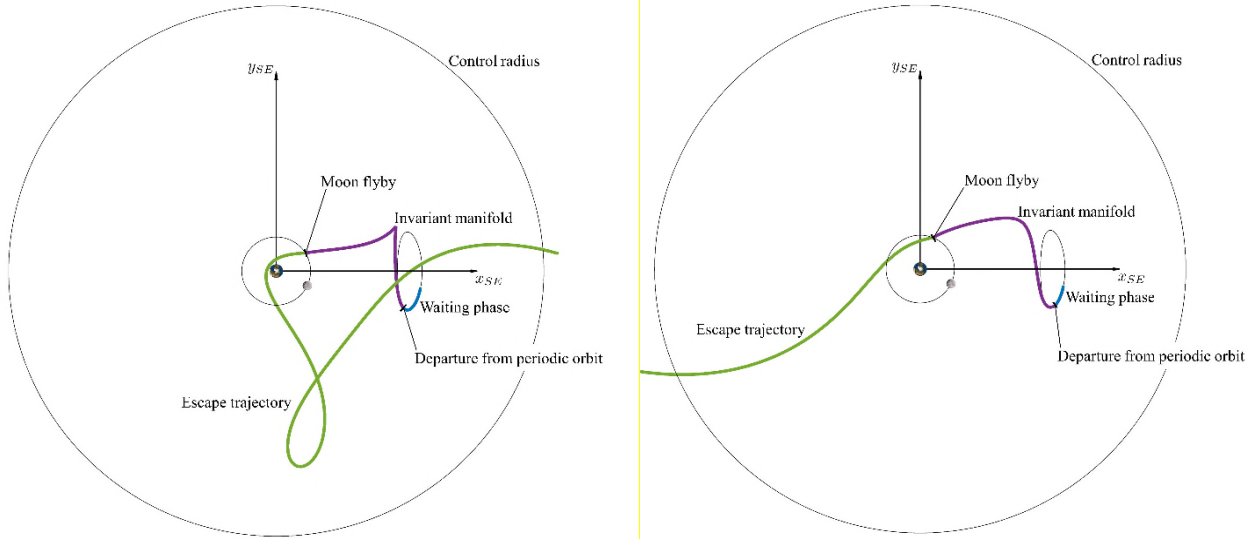

**Figure 15. Two examples of Comet-I departure transfers including a lunar swing-by.** On the right, the departure starts on the negative branch of the invariant manifold, it intersects the Moon and, finally, departs through the L2 direction towards an intersect transfer in an Earth-trailing configuration (due to the semimajor axis larger than 1 au of its heliocentric orbit). On the left, the departure also starts on the negative branch, intersect the Moon and finally departs on the L1 direction: this implies an Earth-leading configuration for its intersection orbit.

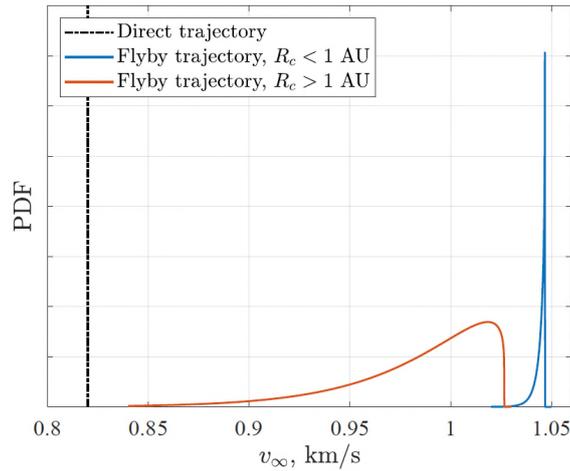

**Figure 16. PDF of the escape velocities obtained from the Monte Carlo Simulation (Ranuschio [17]).**

Figure 16 shows the probability density function (PDF) resulting from the 400 Monte Carlo runs of the aforementioned process [17]. These results demonstrate that the expected benefit in terms of $\Delta v$ of including one single lunar swing-by is of about 200 m/s. Figure 17 instead shows the PDFs for the departure transfer time of flight (*ToF*). This time of flight is defined as the time it takes the trajectory to reach a distance of 0.2 au from the Earth [17]. In order to properly benchmark these results a similar Monte Carlo analysis is generated for the direct departure transfer option (no lunar swing-by). Note that these time of flights to escape are equivalent to those in Section 5.2, yet they are computed up to the time the escape trajectory reaches a distance one order of magnitude larger than in the previous section. As a consequence, the times are also larger as well.

As expected, when Earth-trailing interception are considered (Rc>1 au) the use of the negative branch of the hyperbolic manifold implies longer *ToF* even if a lunar swing-by is included. However, when an Earth-leading interception is attempted (Rc<1 AU), the lunar swing-by has the double benefit of shortening the *ToF* by 200 days, as well as increasing the $v_\infty$-equivalent excess velocity by 200 m/s. Note that the flat homogenous pdf for direct transfers is due to the fact that the upper bound of the waiting time at the LPO is 0.8 years, which is larger than its complete period. As a consequence, all direct transfers simply wait for the optimal conditions to depart.

To conclude, this analysis demonstrates that, regardless of the exact phase of both spacecraft and Moon on their respective orbits at the time of the target LPC discovery, if sufficient warning time is available, a well-designed lunar gravity assist will reduce transfers costs to the comet by about 200 m/s; allowing for a total of 1 km/s of "free" $v_\infty$-equivalent Earth excess velocity. Also, and perhaps more importantly, a lunar swing-by will be beneficious both in terms of extra $\Delta v$-budget and shortening transfer times for interceptions in Earth-leading configurations. It should be noted that there is a 50% chance that the targeted LPC requires such an interception configuration.

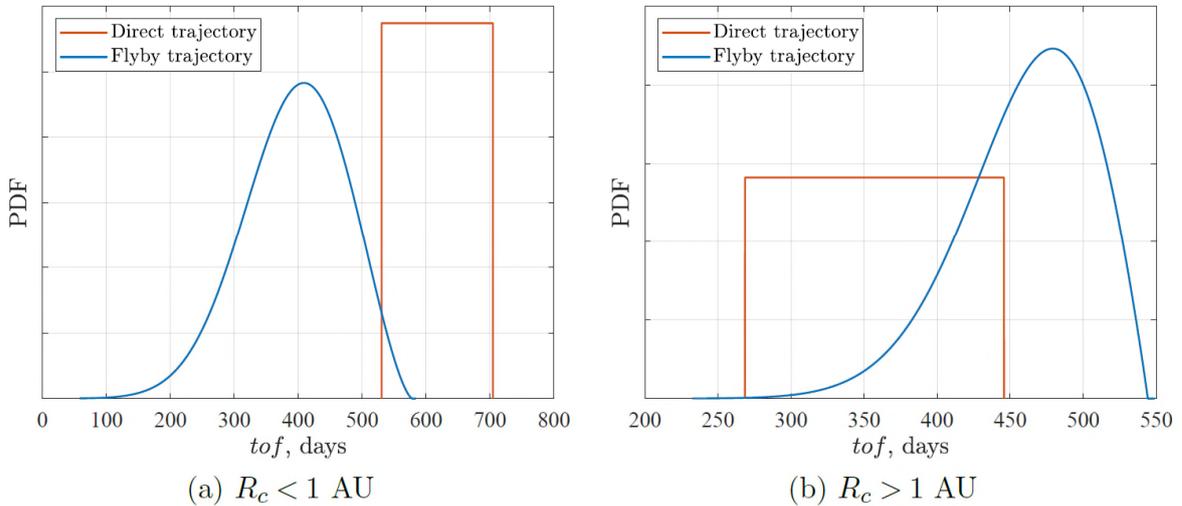

Figure 17. Comparison between the time of flight of the direct escape and the flyby escape obtained from the Monte Carlo Simulation.

### 6. Comet Interceptor Monte Carlo Simulation

Given the uncertainties on the availability of accessible LPC targets, it is key to understand the potential risk that Comet-I may not have the opportunity to fly to a pristine LPC. In order to provide a rough estimate for this risk, as well as to inform the mission design such that the risk may be minimized, a Monte Carlo simulation environment is set up.

This Monte Carlo simulates 10,000 mission scenarios. For each mission scenario, an average of 2.1 LPCs are randomly generated yearly for a mission duration spanning a maximum of 6 years. The decade 2010-2019 saw a total of 21 new LPCs with a nodal point within the accessible range [0.9,1.25] au; i.e. 2.1 LPCs a year. This range of [0.9,1.25] au is used here in order to account for both $\Delta v$ budget constraints, as well as thermal design constraints [3]. The intersection points of each LPC are randomly generated using a homogenous distribution for both heliocentric distance Rc and Earth phase angle $\theta$ of the intersection. Next, a warning time is also allocated to each randomly generated LPC, based on the results of an LPC detection efficiency analysis of the survey simulation environment developed for the LSST [2, 5, 9, 10].

The transfer $\Delta v$ cost to intercept a randomly generated LPC is then interpolated from the data set presented in Section 4. This Monte Carlo Analysis allows extracting a reliable measure of the probability to complete the mission within an allocated mission timespan. The results of this analysis are summarized in Figure 18.

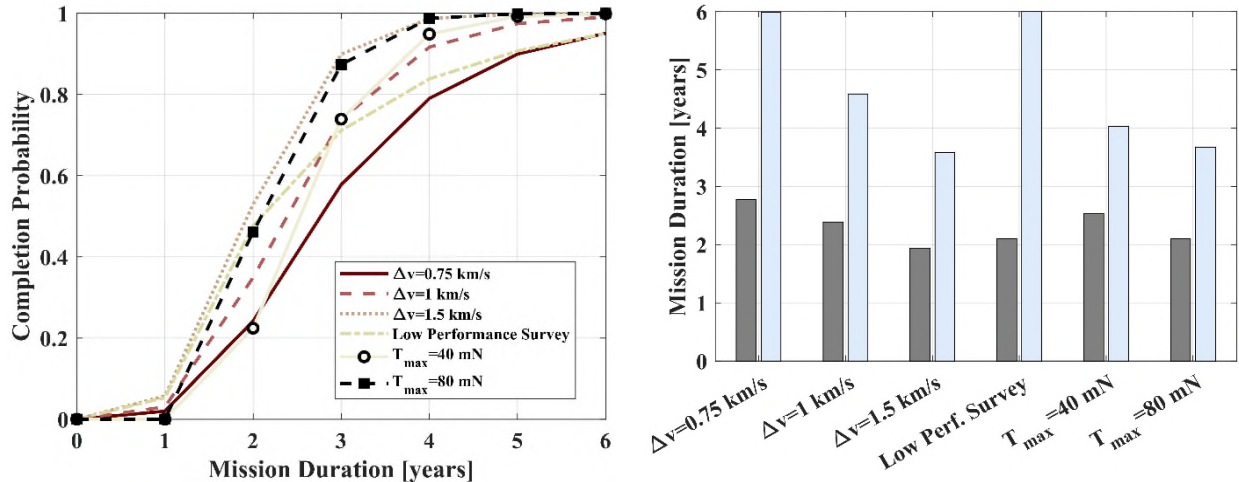

**Figure 18. Probability of successfully completing Comet-I mission within the allocated mission timespan (i.e., 6 year).** Left figure: Cumulative probability as a function of mission duration for a set of different scenarios and $\Delta v$ budgets. Right figure: median mission duration and 95%-confidence mission duration for the same set of mission scenarios. These mission scenarios are: (1) $\Delta v$-budget of 750 m/s, (2) $\Delta v$-budget of 1 km/s, (3) $\Delta v$-budget of 1.5 km/s, (4) $\Delta v$-budget of 1.5 km/s and a survey performance featuring a (40-17-7)% discovery efficiency, rather than the (90-80-75)% used in the rest of the scenarios, (5) Continuous low thrust transfer with maximum thrust capability at 40 mN, (6) Continuous low thrust transfer with maximum thrust capability at 80 mN. Both low thrust scenarios (5 and 6) assume 89 kg of propellant and this an equivalent $\Delta v$ capability of 1.5 km/s.

Before conclusions can be drawn, results in Figure 18 need to be discussed in detail. Firstly, it shall be reminded that, as described in Section 4, the transfer costs assume a patched-conic framework with departure from the SE L2/L1 points. As stated in the Comet-I Mission requirements document [1], Comet-I propulsive capacity will likely be of 750 m/s, if the final design settles for a chemical propulsion configuration [3]. Nevertheless, in a patched-conic framework as described in Section 4, the $\Delta v$ performance includes the entirety of the departure $v_\infty$ as part of its $\Delta v$ costs, while, as discussed in Section 5, an important part of the departure $v_\infty$ can be achieved *for free* by exploiting the gravitational perturbations of the Earth and the Moon. In particular, Figure 15 shows how while direct departure options may provide about 800 m/s of excess velocity, a lunar swing-by may allow about 1 km/s. Hence, a lunar fly-by option is likely to add +700 m/s to the dynamical framework set-up considered in Figure 18 (recall that 300 m/s were gained from a SE L2/L1 point departure [17]). Hence, the effective $\Delta v$ performance of Comet-I should be somewhere in the range of 1 to 1.5 km/s for the purpose of this Monte Carlo results.

Another important source of uncertainty relates with the expected discovery efficiency of the LSST survey, and the associated warning time for the target LPCs. All scenarios generated in Figure 18, except scenario (4), which is labelled as *Low Performance Survey*, consider a warning time (i.e., time from discovery of the LPC to the interception opportunity) following the estimated LSST discovery efficiency reported by Jones et al. [2]. As stated in Section 2, this is equivalent to assume that 90% of the LPCs will be spotted beyond the orbit of Jupiter, 80% should allow for more than 2 years of warning time and about 75% should allow for more than 3 years of warning time. However, there is still substantial an uncertainty on these warning times, which spawn from the unknowns on the evolution of the brightness expected for dynamically new or very *fresh* LPC targets, among other factors. In an attempt to understand the sensitivity of the Monte Carlo results related with the uncertainties on the warning time, a *Low Performance Survey* able to detect only 40% of the LPCs beyond the orbit of Jupiter, 17% with more than 2 years and 7% with more than 3 years warning time is found to achieve a 95% success confidence with 6 years of mission duration. Considering that the LSST survey will scan the night sky at a visual limiting magnitude three points deeper than current surveys, this survey performance seems indeed achievable.

Hence, results in Figure 18 provide only a preliminary estimate and should be followed by a much more specific Monte Carlo simulation. Nevertheless, they should provide a solid confirmation of the robustness of the mission concept. Indeed, the results show how the risk of not having at least one good opportunity to fly-by an LPC is low; in the order of 1 to 5 %. Interestingly, it also shows how there is a good probability that the mission will be completed in very few years. The median mission duration (i.e. 50% completion likelihood) is ranges from 2 to 3 years for all the scenarios represented in Figure 18. Note that the mission duration has not considered neither the time from launch to the decommissioning into the L2 orbit, nor the time from the comet fly-by to the completion of the transmission of all the science data.

Nevertheless, even in the unlikely event that no accessible LPC is found during the allocated waiting time in L2 (~3 to 4 years), Comet-I science team identified a list of potential backup short period comets as alternative targets [12, 20].

# 7. Conclusions.

The lack of specific target and the uncertainties on the orbital characteristics of the future discovered LPCs, implies large mission uncertainties, which bring clear challenges to Comet-I mission design. Hence, the mission design must be underpinned by a solid understanding of the trajectory options and the relative merits of different design alternatives. Given the lack of specific target, these merits can only be measured on a statistical basis, in terms, for example, of the accessible region or the mission success probability.

The trajectory design of Comet-I can be divided in two clear phases; the Earth departure and the heliocentric intercept trajectory. These have been studied separately through three different tightly supervised MSc research projects and their results are synthetized in the document. The different models and conclusions allow to build a preliminary Monte Carlo simulation that demonstrate a 95 to 99% likelihood of successfully visiting a pristine LPC with a chemical propulsion configuration ($\Delta v$~750 m/s) and the added benefit of some third-body gymnastics to depart the Earth's gravity well.

Ultimately, this paper brings to a conclusion the Cranfield-DEIMOS Space research collaboration for Comet-I mission studies that was carried out in the second and third quarters of 2020.

## Acknowledgements

The authors would like to acknowledge the support of ESA and all the national space agencies involved in Comet-I, as well as all the engineers and scientist at ESA and the Science Team that contribute towards a successful mission. Lastly, the authors would also like to thank to all of those who made the work in times of pandemic possible, which include both IT support and our families at home.